\documentclass{nature}
\usepackage{lineno}
\usepackage{amsmath,amssymb,bm,braket,upgreek}
\usepackage{siunitx}
\usepackage{graphicx,xcolor}
\usepackage{booktabs,multirow,array}
\usepackage[hidelinks]{hyperref}

\usepackage[normalem]{ulem}

\def\ket#1{\left| #1 \right\rangle}
\def\bra#1{\left\langle #1 \right|}

\newcommand{\beq}{\begin{equation}}
\newcommand{\eeq}{\end{equation}}

\usepackage{authblk}
\usepackage[english]{babel}
\usepackage{capt-of}

\makeatletter
\let\saved@includegraphics\includegraphics
\AtBeginDocument{\let\includegraphics\saved@includegraphics}
\renewenvironment*{figure}{\@float{figure}}{\end@float}
\makeatother

\title{A building block of quantum repeaters for scalable quantum networks}

\begin{document}

\author[1,2,3]{Wen-Zhao Liu}
\author[1,2,3]{Ya-Bin Zhou}
\author[1,2,3]{Jiu-Peng Chen}
\author[3,4]{Bin Wang}
\author[1,2,3]{Ao Teng}
\author[1,2,3]{Xiao-Wen Han}
\author[1,2,3]{Guang-Cheng Liu}
\author[1,2,3]{Zhi-Jiong Zhang}
\author[1,2,3,4]{Yi Yang}
\author[1,2,3]{Feng-Guang Liu}
\author[1,2,3]{Chao-Hui Xue}
\author[1,2,3]{Bo-Wen Yang}
\author[1,2,3,5]{Jin Yang}
\author[1,2,3]{Chao Zeng}
\author[2]{Du-Ruo Pan}
\author[3,4]{Ming-Yang Zheng}
\author[6]{Xing-Jian Zhang}
\author[6]{Cao Shen}
\author[1,2,3]{Yi-Zheng Zhen}
\author[7]{You Xiao}
\author[7]{Hao Li}
\author[7]{Li-Xing You}
\author[3,8]{Xiong-Feng Ma}
\author[6]{Qi Zhao}
\author[1,2,3]{Feihu Xu}
\author[1,2,3]{Ye Wang}
\author[1,2,3]{Yong Wan}
\author[1,2,3,4]{Qiang Zhang}
\author[1,2,3]{Jian-Wei Pan}

\affil[1]{Hefei National Research Center for Physical Sciences at the Microscale and School of Physical Sciences, University of Science and Technology of China, Hefei, Anhui 230026, China}
\affil[2]{Shanghai Research Center for Quantum Science and CAS Center for Excellence in Quantum Information and Quantum Physics, University of Science and Technology of China, Shanghai, China}
\affil[3]{Hefei National Laboratory, University of Science and Technology of China, Hefei, China}
\affil[4]{Jinan Institute of Quantum Technology, Jinan, China}
\affil[5]{Department of Physics, Anhui Normal University, Wuhu, Anhui 241000, China}
\affil[6]{QICI Quantum Information and Computation Initiative, School of Computing and Data Science, The University of Hong Kong, Pokfulam Road, Hong Kong SAR, China}
\affil[7]{National Key Laboratory of Materials for Integrated Circuits, Shanghai Institute of Microsystem and Information Technology, Chinese Academy of Sciences (SIMIT, CAS)}
\affil[8]{Center for Quantum Information, Institute for Interdisciplinary Information Sciences, Tsinghua University, Beijing, 100084, China}

\maketitle

\newpage
\begin{abstract}
\sloppy{
Quantum networks, integrating quantum communication, quantum metrology, and distributed quantum computing, could provide secure and efficient information transfer, high-resolution sensing, and an exponential speed-up in information processing\cite{kimble2008quantum}.
Deterministic entanglement distribution over long distances is a prerequisite for scalable quantum networks, enabling the utilization of device-independent quantum key distribution (DI-QKD)\cite{bennett1984quantum,ekert1991quantum} and quantum teleportation\cite{bennett1993teleporting,bouwmeester1997experimental} to achieve secure and efficient information transfer. 
However, the exponential photon loss in optical fibres prohibits efficient and deterministic entanglement distribution. 
Quantum repeaters\cite{briegel1998quantum}, incorporating entanglement swapping\cite{bennett1993teleporting,zukowski1993event,pan1998experimental} and entanglement purification\cite{bennett1996purification,deutsch1996quantum,pan2003experimental} with quantum memories, offer the most promising means to overcome this limitation in fibre-based quantum networks.
Despite numerous pioneering efforts toward realizing quantum repeaters\cite{chou2007functional,moehring2007entanglement, yuan2008experimental,hofmann2012heralded,bernien2013heralded,hucul2015modular, delteil2016generation,humphreys2018deterministic,yu2020entanglement,stephenson2020high,van2022entangling,liu2024creation,knaut2024entanglement,stolk2024metropolitan}, a critical bottleneck remains, as remote memory-memory entanglement suffers from decoherence more rapidly than it can be established and purified over long distances.
We overcome this by developing long-lived trapped-ion memories, an efficient telecom interface, and a high-visibility single-photon entanglement protocol\cite{cabrillo1999creation,duan2001long}.
This allows us to establish and maintain memory–memory entanglement over a 10 km fibre within the average entanglement establishment time for the same distance.
As a direct application, we demonstrate metropolitan-scale DI-QKD, distilling 1,917 secret keys out of $4.05\times10^5$ Bell pairs over 10~km. We further report a positive key rate over 101 km in the asymptotic limit, extending the achievable distance by more than two orders of magnitude\cite{nadlinger2022experimental,zhang2022device,liu2022toward}. 
Our work provides a critical building block for quantum repeaters and marks an important step toward scalable quantum networks.
}
\end{abstract}

Quantum repeaters offer the most promising approach to overcoming exponential photon loss in fibre-based quantum networks\cite{briegel1998quantum}.  
Instead of sending photons directly between users, a quantum repeater establishes intermediate nodes and employs entanglement swapping\cite{bennett1993teleporting,zukowski1993event,pan1998experimental}  with quantum memories to mitigate photon loss. 
For instance, direct entanglement distribution over 1,000 km of fibre would suffer from a photon loss of $10^{20}$, whereas introducing 10 repeater nodes could reduce this loss to about $10^2$. 
This approach significantly enhances distribution efficiency and enables deterministic entanglement over long distances.  
As an essential prerequisite for quantum repeaters, quantum entanglement must survive longer than the entanglement establishment time at each node, so neighbouring nodes can be connected via entanglement swapping. 
Moreover, high-fidelity deterministic entanglement enables device-independent quantum key distribution (DI-QKD), an ultimate goal of quantum cryptography, which provides information-theoretic security by certifying secret-key generation through violations of Bell inequalities, regardless of device trustworthiness or channel integrity\cite{ekert1991quantum,mayers1998quantum,zhang2023quantum}. 

Establishing deterministic entanglement over metropolitan distances has been successfully demonstrated across various experimental platforms, including atomic ensembles\cite{yu2020entanglement, liu2024creation}, single atoms\cite{van2022entangling}, and color centres in diamond\cite{stolk2024metropolitan,knaut2024entanglement}.
However, in all these experiments, owing to limited initial fidelity and significant decoherence, the entanglement losts within the establishment time.  
As a result, such entanglement cannot be employed for multi-stage entanglement swapping or purification, thereby preventing the realization of practical quantum repeaters. 

Here we report the experimental realization that overcomes the aforementioned challenges.
We develop long-lived $^{40}$Ca$^+$ ion memories, efficient and low-noise telecom interfaces based on periodically poled lithium niobate (PPLN) waveguides, and a highly-visibility single-photon entanglement protocol\cite{cabrillo1999creation,duan2001long}. 
We establish memory-memory entanglement over 10 km fibre, complemented by near-unity state readout capability.  
The coherence time of memory–memory entanglement is 550 ± 36 ms, exceeding the average entanglement generation time of 450 ms for the same distance. The expected fidelity of a Bell pair stored from a previous entanglement round is 0.578 ± 0.006 at the time a subsequent round succeeds.
This result confirms that multiple memory–memory entanglements can be established and maintained across metropolitan-scale nodes, enabling multi-stage entanglement swapping and entanglement purification, thereby serving as a crucial building block for quantum repeaters in scalable quantum networks.

As a powerful application of our building block, we demonstrate long-distance DI-QKD.  
We establish remote entanglements with a fidelity exceeding $0.90$ over separation of up to 101 km of fibre.  
State measurements are performed immediately after entanglement establishment to test Bell inequalities and generate keys for DI-QKD. 
We distill 1,917 secret keys out of $4.05 \times 10^5$ Bell pairs over a 10-km fibre and obtain a positive key rate over 101 km.

\section*{Heralded entanglement between nodes}
\begin{figure}[t]
\includegraphics[width=1\linewidth]{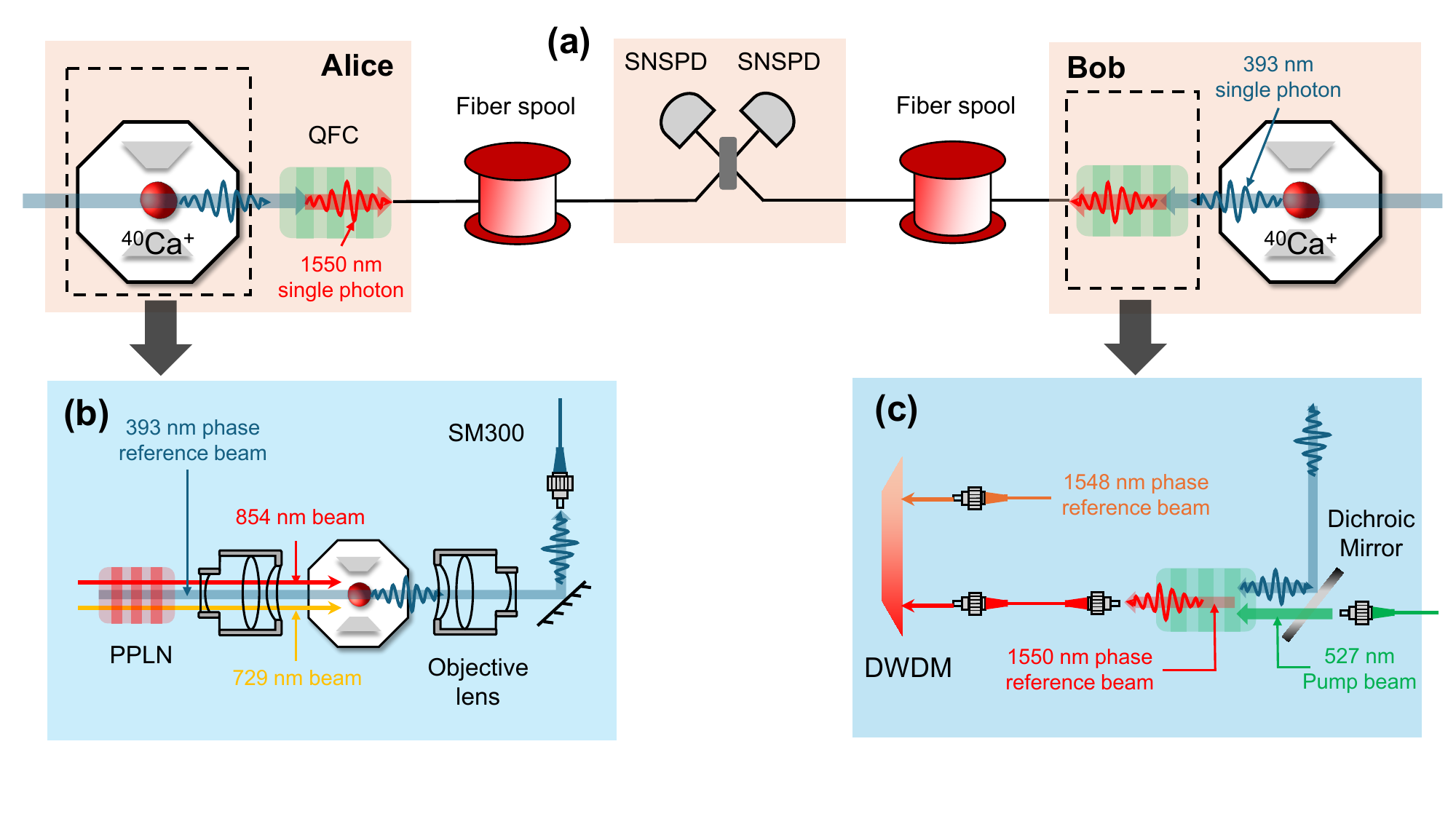}
\captionof{figure}{\textbf{Experimental setup schematics:} 
(a) Overview of Experimental setup. Single photons generated from two independent trapped-ion nodes interfere at a midpoint along the fibre spool. Successful photon detection on either detectors at the middle station heralds ion–ion entanglement. (b) Laser beams at 729 and 854 nm for coherent state manipulation and single photon generation are sent into the trap along the optic axis of objectives. The same beams also generate 393~nm phase reference pulse via sum-frequency generation before the trap. (c) The 393~nm photons are combined with a 527~nm pump laser beam at a dichroic mirror and converted to 1550~nm via waveguide-based difference-frequency generation. A 1548~nm phase reference beam is coupled into the optical link via a dense wavelength-division multiplexer (DWDM) to pre-stabilize the long fibre.}
\label{fig:schematics}
\end{figure}

To establish an elementary link in a quantum network, we generate entanglement between two independent quantum nodes, Alice and Bob, each hosting a single $^{40}$Ca$^+$ ion and connected by telecom-band fibre of varying lengths. 
Single photons emitted by each node are collected using high numerical aperture (NA = 0.635) objective lenses, coupled into optical fibres, sent through a quantum frequency conversion (QFC) module, and transmitted over long fibres to an intermediate node. Entanglement between two nodes is achieved via phase-stabilized single-photon interference and heralded by clicks on the superconducting nanowire single photon detectors (SNSPDs) (Fig.~\ref{fig:schematics}a).

At each node, ions are Doppler cooled, followed by electromagnetically-induced-transparency (EIT) cooling of all motional modes. State initialization into $\ket{S_{1/2}, m_j=+1/2} \equiv \ket{\downarrow}$ is then performed via frequency-selective optical pumping on the $\ket{S_{1/2}, m_j=-1/2} \leftrightarrow \ket{D_{5/2}, m_j=+3/2}$ transition, followed by repumping to the $\ket{S_{1/2}}$ manifold, achieving a typical preparation fidelity of 99.9\%. Qubits are encoded in $\ket{\downarrow}$ and $\ket{D_{5/2}, m_j=+5/2} \equiv \ket{\uparrow}$, and read out via state-dependent fluorescence detection. 

\begin{figure}[t]
\includegraphics[width=1\linewidth]{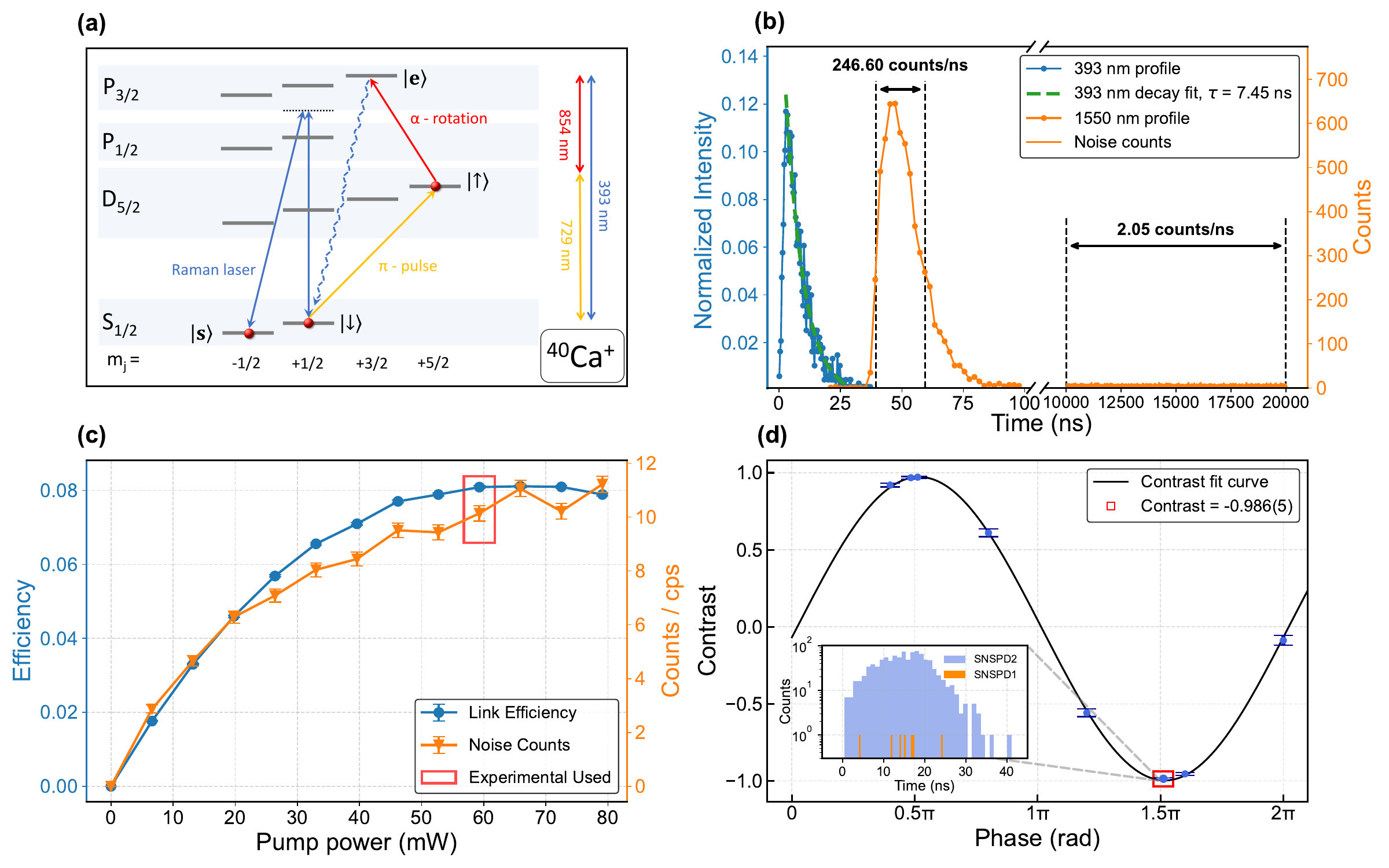}
\captionof{figure}{\textbf{Entanglement generation:}
(a) 
Energy level diagram of a calcium-40 ion, illustrating single-photon generation via the metastable $\ket{\uparrow}$ and the excited states $\ket{e}$, and single qubit rotation between two Zeeman sub-levels $\ket{\downarrow}$ and $\ket{e}$  using Raman transitions. (b) Temporal profile of single photons at 393 and 1550~nm, detected using a PMT and a SNSPD respectively. (c) Efficiency and noise characteristics of the QFC module (393~nm$\rightarrow$1550~nm) as a function of pump power at 527~nm after transmission through a 10~km fibre link. (d) Contrast measurement for the phase reference laser, with the corresponding standard deviation shown. The inset shows the histogram of detector counts at the point of maximum contrast.}
\label{fig:exp-setup}
\end{figure}

Following initialization, single photons are generated in two steps: (i) a $\pi$-rotation on the 729~nm transition transfers the population from $\ket{\downarrow}$ to $\ket{\uparrow}$, and (ii) a 2.25-ns pulse at 854~nm transfers a fractional population, $\alpha$, from $\ket{\uparrow}$ to $\ket{\downarrow}$ via the intermediate state $\ket{P_{3/2}, m_j=3/2}\equiv\ket{e}$ (Fig.~\ref{fig:exp-setup}a). The resulting 393\,nm photons are collected with custom-designed objectives oriented perpendicular to the quantization axis, yielding an estimated $9\%$ collection efficiency for photons emitted on the $\sigma$-transition.
After coupling into single-mode fibres (SM300), we measure single photon efficiencies of 2.6\% for Alice and 2.8\% for Bob prior to the QFC module. The temporal profile of the 393~nm photons, recorded with a photomultiplier tube, reflects the spontaneous decay lifetime of the excited state $\ket{e}$. The protocol generates entanglement between ions' internal states and photon number states.

Single photons at 393~nm are injected into a frequency-conversion module comprising of a PPLN waveguide and a 527~nm pump laser, enabling single-step conversion to 1550~nm (Fig.~\ref{fig:schematics}c). The broadband noise intrinsic to the conversion process reaches $1.3\times10^{5}$ photons s$^{-1}$ nm$^{-1}$ around 1550~nm at 50~mW of pump power. At the same pump power, the noise is reduced to 35~cps using a 100 GHz bandpass filter, a 10 GHz volume Bragg grating (VBG), and a 40 MHz etalon, while maintaining a combined transmission efficiency of 28\%. The 1550~nm photons, exhibit temporal broadening with a full width at half maximum of about 20 ns due to ring-down in the etalon (Fig.~\ref{fig:exp-setup}b).

To enhance the entanglement generation rate between spatially separated ion qubits, we employ the SPEP. This requires stable phase difference between the optical fields from Alice and Bob\cite{humphreys2018deterministic,yu2020entanglement}. Alice and Bob share a common laser system, but are controlled with independent electronics. Phase fluctuations from both the source and the optical paths are compensated by a stabilization scheme combining wavelength-division multiplexing (WDM) and time-division multiplexing (TDM).

In the WDM, a continuous-wave 1548~nm reference laser is multiplexed with the 1550~nm signal photons immediately after the QFC module (Fig.~\ref{fig:schematics}c), allowing active suppression of phase noise in the fibre links. The TDM scheme compensates slow drifts using a weak 393\,nm reference pulse. This pulse is generated via sum-frequency mixing of the same 729\,nm and 854\,nm beams previously employed for single-photon generation (Fig.~\ref{fig:schematics}b).
The weak reference pulse propagates with the signal photons along a nearly identical path and is applied only during Doppler cooling to avoid heating of the ions’ motional states. With both stabilization schemes implemented, an interference contrast of $0.986 \pm 0.005$ is achieved at the intermediate node from two reference pulses transmitted through a 10 km fibre link (Fig.~\ref{fig:exp-setup}d).

\section*{Characterization of Entanglement}
\begin{figure}[t]
\includegraphics[width=1.0\linewidth]{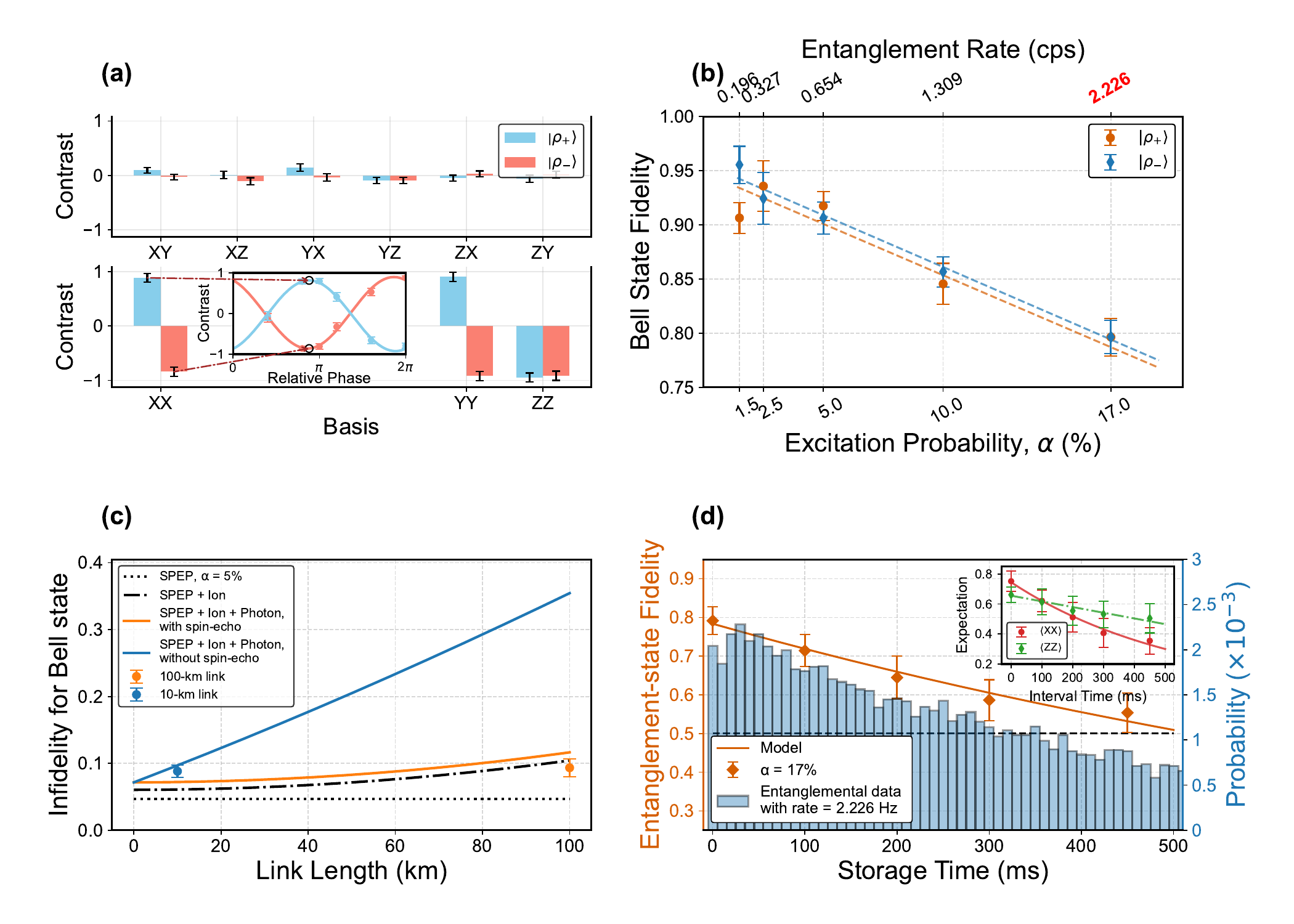}
\captionof{figure}{\textbf{Entanglement fidelity:}
(a) Quantum state tomography and $\braket{XX}$-parity analysis at a fibre length of 10 km, with error bars indicating the standard deviation. $\mathrm{X}$, $\mathrm{Y}$, and $\mathrm{Z}$ denote the Pauli operators $\sigma_x$, $\sigma_y$, and $\sigma_z$, respectively. (b) Fidelity of the $\rho_{\pm}$ states, with error bars indicating the standard deviation, extracted from the population and parity measurement as a function of the excitation parameter $\alpha$ and entanglement generation rate. (c) Infidelity as a function of fibre length, with error bars denoting the standard deviation. Contributions are attributed to protocol-induced error (1), ion-related effects (2, 5), and photon-related factors (3, 4), following the numbering defined in the main text. (d) The entanglement fidelities (orange squares) are shown for different storage times, while the corresponding expectation values of $\braket{XX}$ (red circles)  and $\braket{ZZ}$ (green diamonds) are shown in the inset. The error bars representing one standard deviation. The probability distribution of intervals between two consecutive entanglement events is plotted with a bin size of 10~ms.}
\label{fig:heralded_entanglement} 
\end{figure}

The SPEP produces heralded entanglement between Alice and Bob in the form
\begin{equation}
\ket{\rho_\mathrm{AB}} = (1-\alpha)\ket{\psi^{\pm}}\bra{\psi^{\pm}}+\alpha\ket{\downarrow\downarrow}\bra{\downarrow\downarrow},
\end{equation}
where $\ket{\psi^{\pm}} = (\ket{\downarrow\uparrow} \pm e^{i\Delta \phi }\ket{\uparrow\downarrow})/\sqrt{2}$ carries a differential phase $\Delta\phi$ between optical fields from two nodes. 
The dominant error mechanisms in heralded ion–ion entanglement include: (1) simultaneous excitation at both nodes intrinsic to the SPEP, (2) operational and decoherence errors of the ions, (3) spurious heralding signals from QFC and detector noise, (4) residual phase fluctuations of the cross-node optical fields, and (5) Motional degrees of freedom of the ions, where the excitation–emission cycle imparts a net momentum kick, producing spin–motion entanglement that reduces interference visibility \cite{eschnerLightInterferenceSingle2001,slodickaAtomatomEntanglementSinglephoton2013}. The first four contributions are quantified using experimental data from individual submodules (Supplementary Information), and all errors are modeled shown in Fig.~\ref{fig:heralded_entanglement}c. Notably, there exists a trade-off between protocol-induced errors, governed by $\alpha$, and the entanglement generation rate. At $\alpha = 17\%$, the rate reaches 2.226~cps, corresponding to an average generation time of 450~ms (Fig.~\ref{fig:heralded_entanglement}b).


For a 10-km fibre link, a 527-nm pump power of $\sim$60~mW yields a link efficiency of 9.1\% at $\alpha = 2.5\%$, with a noise level of 9.6~cps, corresponding to a signal-to-noise ratio exceeding 100:1 (Fig.~\ref{fig:exp-setup}b,c). We perform quantum state tomography on the heralded states $\rho_{\pm}(\alpha=2.5\%)$, and determine a state fidelity of $F_{\rho_{+}}=0.923\pm0.012$ and $F_{\rho_{-}}=0.910\pm0.012$ compared to the ideal Bell states $\ket{\Psi^+}$ and $\ket{\Psi^-}$, respectively (Fig.~\ref{fig:heralded_entanglement}a). A simplified estimate using parity and population measurements produces consistent results.  We performed a distance-dependent measurement of entanglement fidelity using 10- and 100-km fibre spools. At a fixed excitation rate of $\alpha = 5\%$, the Bell-state fidelity—determined from population and parity measurements—remains essentially constant, enabled by spin-echo dynamical decoupling that mitigates magnetic-field noise (Fig.~\ref{fig:heralded_entanglement}c).


The lifetime of the remote entanglement is primarily limited by the natural lifetime of the $\ket{D_{5/2}}$ state, which is 1.16 seconds, and by decoherence arising from magnetic-field noise. 
A state $\ket{S_{1/2}, m_j=-1/2}\equiv \ket{s}$ is used to coherently transfer the entanglement state from $(\ket{\downarrow\uparrow} \pm e^{i\Delta \phi}\ket{\uparrow\downarrow})/\sqrt{2}$ to $(\ket{\downarrow s} \pm e^{i\Delta \phi}\ket{s \downarrow})/\sqrt{2}$.
The transfer is achieved in two steps: a stimulated Raman transition between $\ket{s}$ and $\ket{\downarrow}$ is driven by a bi-chromatic 393\,nm Raman beams; a $\pi$-rotation between $\ket{\downarrow}$ and $\ket{\uparrow}$ is applied using a 729\,nm laser at each node.
A subsequent implementation of Knill dynamical decoupling (KDD) pulses\cite{souza2011robust} with an interval of 0.5 ms is applied to mitigate decoherence. 

 
A heralded remote entanglement is generated with an excitation probability of $\alpha = 17\%$ across a 10~km fibre link and is then protected by the storage protocol described above. As shown in Fig.~\ref{fig:heralded_entanglement}d, the entanglement coherence, quantified by the expectation value of $\braket{XX}$, decays with increasing wait time and is well described by an exponential fit. 
The fit yields a coherence time of $550 \pm 36$ ms, corresponding to a decoherence rate of $1.8 \pm 0.1$ Hz. The quantum link efficiency, defined as the ratio between the entanglement generation rate and the decoherence rate, is calculated to be 1.2, exceeding the critical threshold of 0.83 required for the deterministic delivery of remote entanglement\cite{humphreys2018deterministic}.
A linear decay is observed in the expectation value of $\braket{ZZ}$, which we attribute to the accumulation of gate errors during the repeated $\pi$-rotations of the KDD sequence.
Independent measurements show that the gate errors are dominated by spontaneous emission, with per-gate error probabilities
of $6.1\times10^{-4} $ at the Alice node and $5.7\times10^{-4}$ at the Bob node. 
From the measured expectation values of $\braket{XX}$ and $\braket{ZZ}$, the entanglement fidelity is determined to be $0.554 \pm 0.050$ at 450~ms, corresponding to the average generation time at $\alpha = 17\%$. The entanglement remains protected, maintaining a fidelity above 0.5 for up to $547 \pm 34$~ms (Fig.~\ref{fig:heralded_entanglement}(d)).
Further details are provided in the Supplementary Information.

The pre-established entanglement only needs to be preserved until the subsequent entanglement is generated, such that two entangled pairs can then be exploited for swapping or purification.
Accordingly, the average fidelity is evaluated as the probability-weighted mean of the fidelities associated with different time intervals between consecutive entanglement events. 
By fitting the expectation values of $\braket{XX}$ and $\braket{ZZ}$, together with the knowledge of the probability density function of intervals, we can extract the average fidelity. 
When the integration window is restricted to 450 ms—the mean entanglement generation time—the average fidelity is found to be 
$0.668\pm0.005$. Extending the integration window to infinity yields an average fidelity of $0.578\pm0.006$, which represents the expected fidelity of a Bell pair stored from a previous entanglement round once the subsequent round succeeds.

\section*{DI-QKD protocol and results} 
\begin{figure}[t]
\centering
\includegraphics[width=1.0\linewidth]{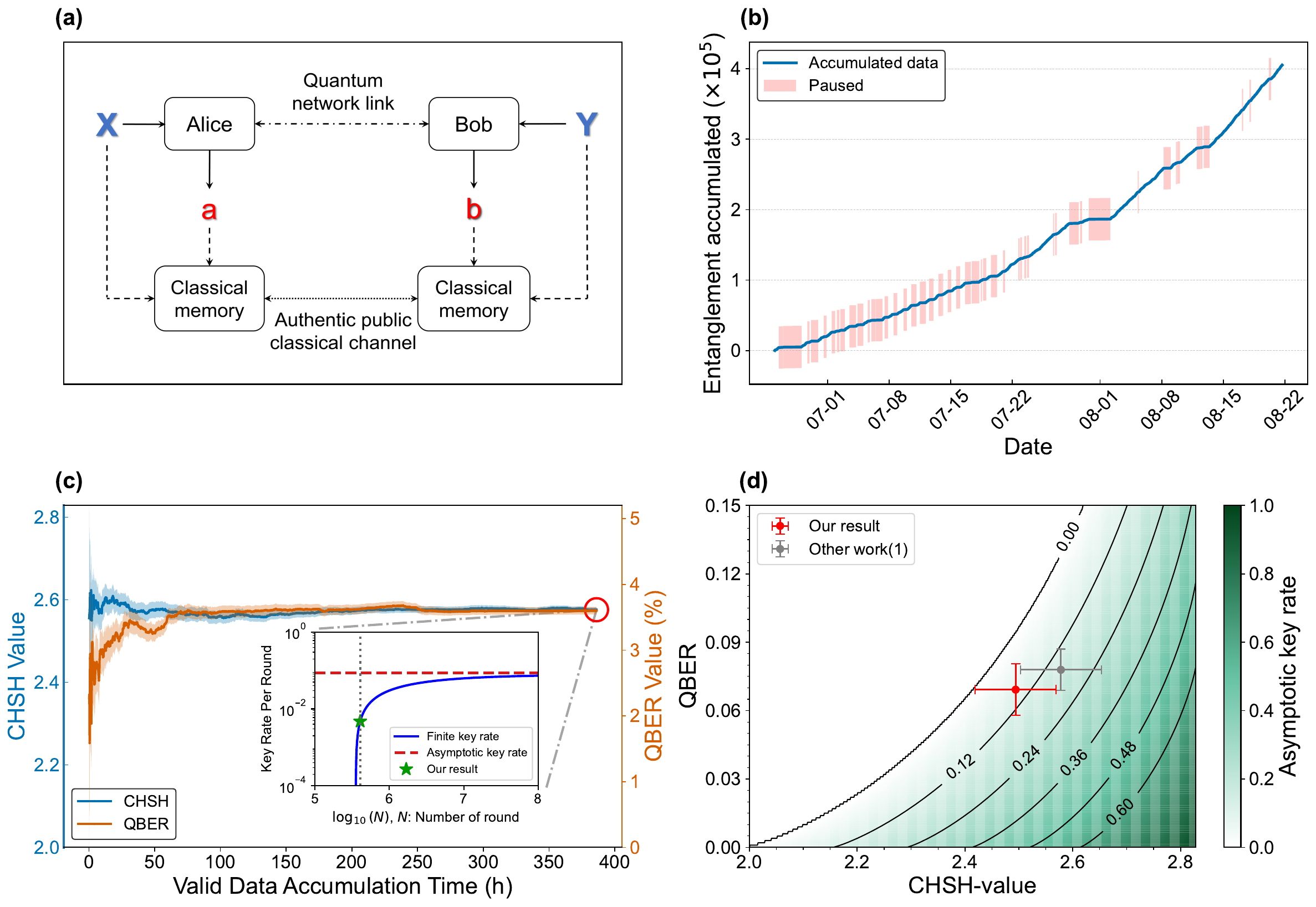}
\captionof{figure}{\textbf{DI-QKD results:} (a) Alice and Bob are connected via a quantum network link. Each party receives inputs $\mathrm{X}$ and $\mathrm{Y}$ and generates outputs $\mathrm{a}$ and $\mathrm{b}$, respectively. The results are stored locally and communicated via an authenticated public classical channel for post-processing. (b) Blue curve: data accumulation over time; red regions: pauses for system recovery. (c) Accumulated CHSH (blue) and QBER (orange) values, with one-standard-deviation uncertainties shown as shaded regions. Inset: finite-key DI-QKD over a 10-km fibre link with total failure probability $\epsilon = 10^{-5}$. (d) Infinite-key analysis of DI-QKD over a 101-km fibre. Results are shown in the limit where the fraction of key-generation rounds approaches unity, with error bars representing one standard deviation for the CHSH value and QBER, compared with Ref.\cite{zhang2022device}.}

\label{fig:data}
\end{figure}

As an application of the high-fidelity, high-rate deterministic entanglement generated in this setup, we demonstrate DI-QKD over metropolitan-scale fibres.  
The DI-QKD architecture used in our experiment is illustrated in Fig.~\ref{fig:data}a. After heralded entanglement generation, the measurement settings are determined by the inputs $\mathrm{X}$ and $\mathrm{Y}$, producing outputs $a$ and $b$ for Alice and Bob, respectively. We implement the DI-QKD protocol proposed by Zhang et al. \cite{zhang2023quantum}, which employs random quantum measurements for both key generation and the Bell test, and applies the corresponding security analysis. Random numbers generated from our previous device-independent quantum randomness experiments\cite{liu2021device} are used to select the measurement bases. Alice and Bob independently choose their settings at random: $\mathrm{X},\mathrm{Y}\in\{0,1\}$ define Bell-test rounds, while $\mathrm{X}=0$ and $\mathrm{Y}=2$ define key-generation rounds. Finally, all inputs and outcomes are recorded in independent, trusted local classical memories.



We define a heralded entanglement generation as an experimental round and performed a total of $N = 405{,}145$ rounds over a 10\,km fibre link, achieving secure key generation after approximately two months of data collection (Fig.~\ref{fig:data}b).
The Clauser–Horne–Shimony–Holt (CHSH) inequality~\cite{clauser1969proposed} is used to evaluate Bell-test rounds, yielding $S = 2.5758\pm0.0059$, while the quantum bit error rate (QBER) extracted from key-generation rounds is $Q = 0.0360\pm0.0006$. The experiment achieved an average entanglement generation rate of 0.291~s$^{-1}$ over a valid accumulation time of 386.1 hours. After information reconciliation with efficiency $f = 1.122$ and subsequent privacy amplification, the accumulated CHSH and QBER values correspond to 1,917 secret bits, yielding a key rate of approximately $4\times10^{-3}$ per round, which asymptotically approaches 0.325 per round with increasing number of entanglement generation events (Fig.~\ref{fig:data}c). For a 100-km implementation, we accumulated $N = 2{,}799$ rounds in a similar manner and obtained $S = 2.504\pm0.075$ and $Q = 0.069\pm0.011$, with the asymptotic key rate of $0.0974$ per round estimated in the limit of infinitely many key-generation rounds, where their fraction approaches unity (Fig.~\ref{fig:data}d, Supplementary Information).


\section*{Summary and Outlook}
In our experiment, we demonstrate high-fidelity deterministic entanglement establishment between two long-lived trapped ion memories over long distances. 
Over a 10 km fibre link, the remote entanglement maintains survive, 
with the expected fidelity of $0.578\pm0.006$ at the time a subsequent entanglement establishment, which is a crucial prerequisite for multi-stage entanglement swapping and entanglement purification over long distances. 
Moreover, using this high-fidelity long-distance deterministic entanglement, together with 400 hours of continuous data collection, we demonstrate finite-size DI-QKD over a 10 km link, extending the secure distance by a factor of 5,000 compared with previous demonstrations.\cite{nadlinger2022experimental}
Additionally, we demonstrate a positive key rate over 101 km in the asymptotic limit, extending the achievable distance by more than two orders of magnitude\cite{zhang2022device,liu2022toward}.

To realize deterministic entanglement establishment over thousands of kilometres, we need to extend the coherence time of the quantum memory, increase the entanglement generation rate, and develop multi-node network infrastructure.  
To further extend the coherence time of the quantum memory, one can introduce an ion species with clock transitions to serve as a storage qubit\cite{drmotaRobustQuantumMemory2023a} or implement a multi-ion memory with coherence encoded in a decoherence-free subspace\cite{kielpinski2001decoherence} .
To increase the remote entangling rate, we can use optical-cavity enhanced photon collection\cite{krutyanskiy2023entanglement}, or trap more ions in each node using reconfigurable linear chains\cite{canteri2025photon} with individually addressed lasers, which would allow linear enhancement of heralding probability via multiplexing. 
To develop multi-node network infrastructure, cross-node optical frequency and phase referencing between independent laser systems are essential. In the present implementation, slow drift compensation across multiple wavelengths remains unavoidable; a frequency comb at each node for both local and remote optical phase referencing will be an integral component of such a field demonstration.

Deterministic entanglement over thousands of kilometres would enable DI-QKD and the transmission of arbitrary quantum states across ultra-long distances, thereby enabling secure and efficient inter-city quantum information transfer and supporting the development of nationwide quantum networks.



\bibliographystyle{naturemag}
\bibliography{bibDIQKD}

\end{document}